\documentclass[letterpaper, 10 pt, conference]{ieeeconf}  
\IEEEoverridecommandlockouts                              
\usepackage{graphics}
\usepackage{epsfig} 
\usepackage{textcomp}
\usepackage{times} 
\usepackage{amsmath} 
\usepackage{mathtools}
\usepackage{amssymb}  
\usepackage{url}
\usepackage{commath}
\usepackage{graphicx}
\usepackage{caption}
\usepackage{multirow}
\usepackage{subcaption}
\usepackage{array} 
\usepackage{url}

\usepackage{matlab-prettifier}
\usepackage{listings}
\usepackage{footnote}
\makesavenoteenv{tabular}
\makesavenoteenv{table}
\makesavenoteenv{figure}

\usepackage{hyperref}

\usepackage{siunitx}

\title{\LARGE \bf
ROSbag-based Multimodal Affective Dataset for Emotional and Cognitive States}

\author{Wonse Jo, Shyam Sundar Kannan, Go-Eum Cha, Ahreum Lee, and Byung-Cheol Min
\thanks{Wonse Jo, Shyam Sundar Kannan, Go-Eum Cha, Ahreum Lee, and Byung-Cheol Min are with SMART Lab, Department of Computer and Information Technology, Purdue University, West Lafayette, IN 47907, USA \tt\small{jow@purdue.edu, kannan9@purdue.edu, cha20@purdue.edu, lahreum@purdue.edu, minb@purdue.edu}}%
}

\begin{document}
\maketitle
\thispagestyle{empty}
\pagestyle{empty}

\begin{abstract}
This paper introduces a new ROSbag-based multimodal affective dataset for emotional and cognitive states generated using the Robot Operating System (ROS). We utilized images and sounds from the International Affective Pictures System (IAPS) and the International Affective Digitized Sounds (IADS) to stimulate targeted emotions (happiness, sadness, anger, fear, surprise, disgust, and neutral), and a dual $N$-back game to stimulate different levels of cognitive workload. 30 human subjects participated in the user study; their physiological data were collected using the latest commercial wearable sensors, behavioral data were collected using hardware devices such as cameras, and subjective assessments were carried out through questionnaires. All data were stored in single ROSbag files rather than in conventional Comma-Separated Values (CSV) files. This not only ensures synchronization of signals and videos in a data set, but also allows researchers to easily analyze and verify their algorithms by connecting directly to this dataset through ROS. The generated affective dataset consists of 1,602 ROSbag files, and the size of the dataset is about 787GB. The dataset is made publicly available. We expect that our dataset can be a great resource for many researchers in the fields of affective computing, Human-Computer Interaction (HCI), and Human-Robot Interaction (HRI).
\end{abstract}

\section{INTRODUCTION}   
\label{sec:introduction}
The recent advancements in wearable devices have increased the attention to affective computing and Human-Computer Interaction (HCI). The easy availability of the wearable sensors has allowed for its integration with affective computing and has given rise to intelligent computing devices that can interpret the affective state of users and provide adaptive feedback to them accordingly. 
For instance, in an autonomous car, the level of autonomy could be dynamically adjusted based on the affective state of the human operator \cite{kowalczuk2019emotion}. 
In addition to the field of HCI, the affective computing has been deeply influencing the field of robotics too, especially Human-Robot Interaction (HRI). For example, in the social robot interaction system, physical conditions of users extracted from cameras (e.g., facial expression and body gestures) and/or physiological states of users collected from sensors used to flexibly change communication methods to reduce human's antipathy toward the robotics system \cite{bera2017sociosense,bera2019emotionally}. 

With the advent of wireless wearable sensors and other commercially available devices like a smartwatch, there has been an increasing interest in estimating human's state from monitoring physiological signals. 
In response to this current trend of monitoring human state using wearable sensors, it is becoming more important to build more physiological datasets based on wearable sensors. 

Moreover, the development of affective state prediction algorithms and estimation methods using machine learning and neural networks has boosted the availability of publically available annotated affective datasets \cite{tzirakis2017end}. The datasets have focused on recording the physiological responses of the participants using various stimuli. 
However, in most of the existing datasets, the data were recorded using laboratory type monitoring devices which are using wired technologies, so caused inconvenience for participants' movement \cite{koelstra2011deap, abadi2015decaf}. 

In addition to the physiological sensor dataset, external behavioral information of the human is also useful in the estimation of the affective state \cite{Shan2012}. 
For example, the facial data are mostly used in affective datasets alongside the physiological sensor data \cite{correa2018amigos, subramanian2016ascertain}. Another external modality that is widely used is the body gesture data \cite{noroozi2018survey}.
However, there are not many studies considering the relationship between physiological signals and behavioral information, so there are not many datasets including both the physiological data and human behavioral data. Therefore, it is necessary to build multimodal datasets that consist of both physiological and behavioral data.

\begin{figure}[t]
    \centering 
		\includegraphics[width=1\linewidth]{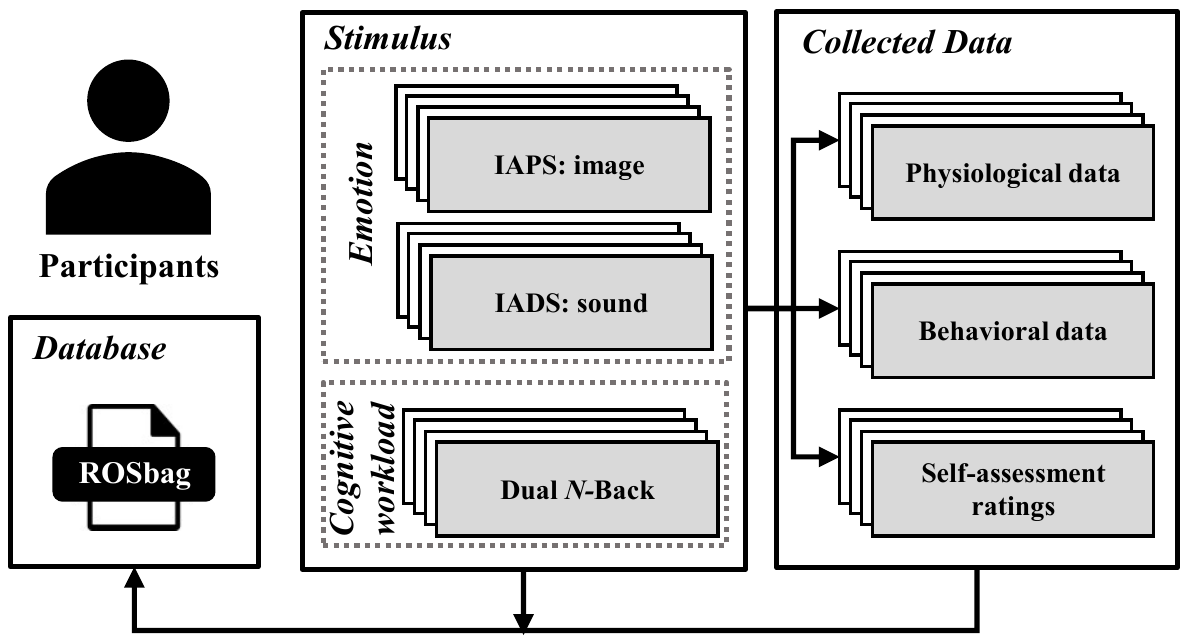} 
    \caption{Outline showing how a new ROSbag-based multimodal affective dataset is created and organized.}
    \label{img:diagram_user_study}
\end{figure}

Furthermore, the estimation of human's affective state for effective HRI has been gaining increased interest in the recent days. The emergence of new robotics middleware (such as Robot Operating System (ROS) \cite{quigley2009ros}) has also played a larger role in growing the variety of HRI research to integrate the robotics system with the affective computing. 
In ROS, the data collected are usually stored as a ROSbag. The ROSbag format has more benefits than the CSV format for collecting and analyzing the dataset. 
Since the ROS can ensure to synchronize the recording signals and videos, it is available to easily and directly analyze the dataset by replaying both using a single ROSbag file. Also, the ROS supports various program languages and operating systems, so that users can validate the developing algorithm and programs by connecting the dataset as like in real-time experiments. Plus, the dataset is available to convert to CSV format or others via additional ROS packages. Therefore, a dataset that combines both physiological and behavioral data based on ROS can have great advantages. 

In this work, we present a ROSbag-based multimodal dataset comprising physiological data measured using wearable devices and behavioral data recorded using external devices. The data were collected from participants through a user study where various stimuli such as images, audio, and workload tasks were used. Fig. \ref{img:diagram_user_study} outlines how the dataset was created and organized. During the user study, physiological responses such as Blood Volume Pulse (BVP), Electrocardiography (ECG), Electro-dermal Activity (EDA), Electromyography (EMG), Galvanic Skin Response (GSR), Heart Rate (HR), Interbeat Interval (IBI), Photoplethysmography (PPG), and Skin Temperature (ST) were measured using commercially available wearable devices. In addition to the physiological sensors, a 3D frontal camera and a side-view camera were used to record face and body gestures, respectively. 
To investigate the implicit behaviors of users, the variations in the keyboard typing and the mouse motion patterns were also recorded. 
During the study, the participants performed a self-assessment of their affective level using questionnaires at the end of each experiment. These subjective data  can be used later for the training of the classifiers.

\section{Related Works} 
\label{sec:related_works}
Human affects shape a huge part of the human experience such as attention, learning, memory, and even decision-making which are required to complete tasks. 
Therefore, understanding and measuring human affects in real-time is vital to construct adaptive and context-aware interfaces that could enrich the user experience.
To do so, affective computing research investigates how affect sensing and elicitation techniques can build the understanding of affect and contribute to the design of technologies \cite{santos2016emotions}. 
Two main methods have been used to estimate human emotion and cognition states \cite{shu2018review}. The first is to analyze internal human changes by monitoring physiological signals such as ECG, GSR, EMG, and so on. 
The other method involves human physical signals such as facial expression, gesture, voice, and so on. 
As human affects are too complex to present with a single signal, many researchers have applied multiple sensors to improve accuracy and reliability of the system \cite{shu2018review,poria2017review}. 

Most affective computing applications use annotated datasets to train machine learning models that recognize human psychological states \cite{poria2017review,mollahosseini2017affectnet}. The majority of the dataset includes multimodal stimuli which were designed to elicit a particular human affect and sensor data that were collected when a subject was exposed to the stimuli. Depending on how the researchers defined the human affects and what types of sensors they used, characteristics of the annotated datasets are different. Although the independence between emotion and cognition is still a controversial topic \cite{leventhal1987relationship}, the researchers mainly focused on emotion recognition by providing different dimensions of emotion, so the affective dataset are getting increasingly diversified (such as, DEAP \cite{koelstra2011deap}, DECAF \cite{abadi2015decaf}, AMIGO \cite{correa2018amigos}, WESAD \cite{schmidt2018introducing}, and so on). 
Most of the existing dataset particularly focused on emotion recognition but did not design a deliberate experimental setting to detect one's cognitive state which could affect one's emotional states. 

In this regard, we present a dataset for detecting emotional and cognitive states which is collected from various wearable devices that can monitor and collect human physiological and behavioral data in an unobtrusive manner. 



\begin{figure*}[t]
    \centering
		\includegraphics[width=0.85\linewidth]{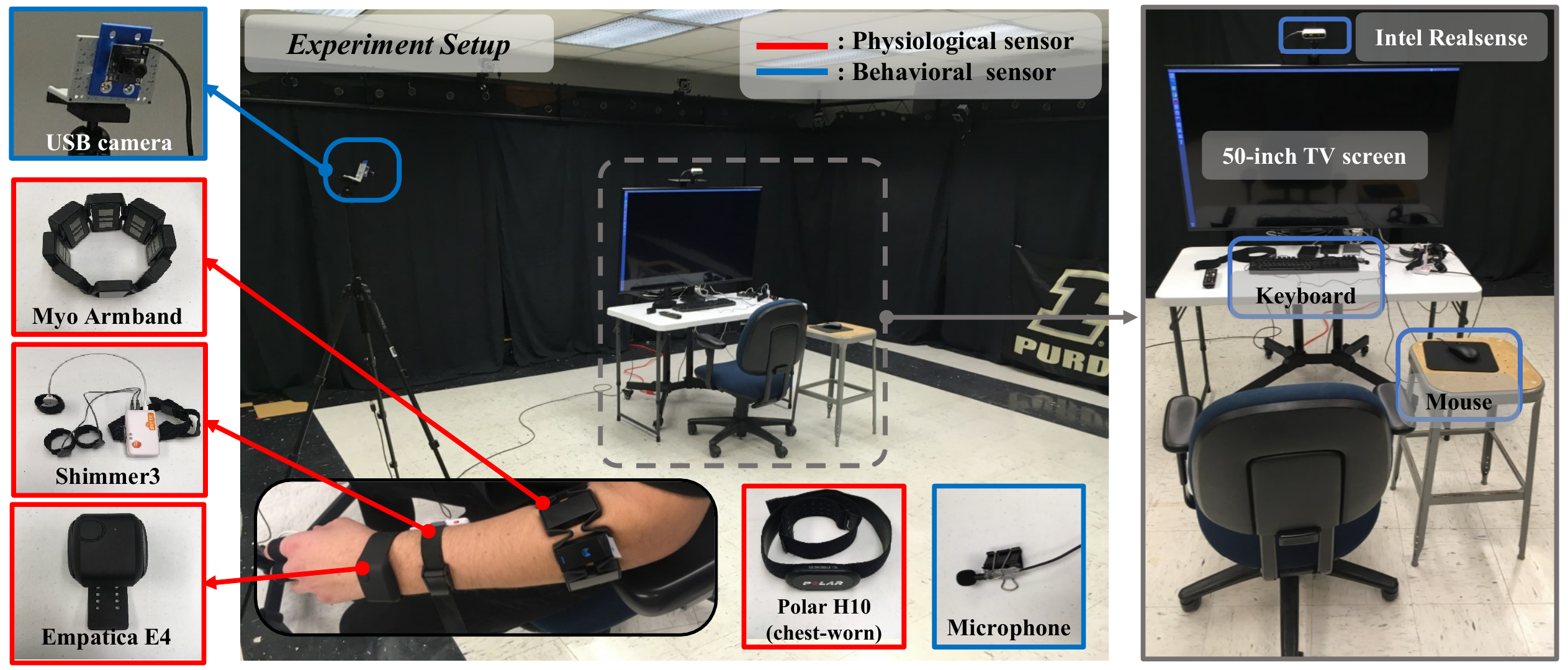} 
    \caption{A user study setting. Commercial wearable sensors including Empatica E4, Shimmer3 GSR and PPG, Polar H10, and Myo armband are utilized. Behavioral sensors including a USB camera (for side view), Intel RealSense (frontal depth and RGB images), a microphone, and a mouse \& keyboard are also utilized.}
    \label{img:experiment_setup}
\end{figure*}


\section{Design of User Study} 
\label{sec:user_study_design}
We designed a user study to build a new affective dataset that includes physiological and behavioral data based on the participants' emotional and cognitive states. All participants were asked to perform two tasks: an emotion elicitation task and a cognitive workload task. This study was approved by the Purdue University’s Institutional Review Board (Purdue IRB Protocol: \#1812021453).  

\subsection{Experimental Setup}
The user study was conducted in a closed indoor setup as shown in Fig. \ref{img:experiment_setup}. The participants were seated in front of a screen with the various wearable sensors and other external sensors connected to a ROS-based monitoring system. Fig. \ref{img:diagram_user_study} depicts the schematic of the monitoring system for reading physiological and behavioral data, as well as self-assessment ratings. The main laptop behind the screen is used to connect all sensors and devices, as well as to execute the Graphical user interface (GUI) programs for displaying emotion stimulus sets and the memory test game on the screen. The programs are connected with the ROS to synchronize and to save the data to a ROSbag file that is used to track and record all rostopic messages communicated within the ROS. 

\subsection{Participants} 
For this user study, we recruited 30 participants from the University; the 11 females and 19 males had an age range of 18 to 37 years (mean: 25.1; std: 4.497). It was ensured that none of the participants had any skin allergies to metal or plastic, medical history of brain disorder, or heart diseases and vision or muscle impairment, so that all the wearable devices could be used. The participants were compensated with \$10 for their participation.

\subsection{Equipment}
\label{sec:data_collection_method}
As shown in Fig. \ref{img:experiment_setup}, the physiological and behavioral sensors used in the monitoring system are wearable and commercial devices, so that the experimental settings do not limit the participant's native body movements which are essential for monitoring. 

The physiological sensors connected to the monitoring system are as follows:
\begin{itemize}
    \item \textbf{Empatica E4} is a wristband with an array of sensors for physiological monitoring: EDA, BVP, IBI HR, and ST \cite{empatica}.
    \item \textbf{Myo} is an armband that measures the 8-channel EMG signals. It includes the 8 electrodes placed inside the band to measure the 8-channels EMG signals \cite{myo_armband}.
    \item \textbf{Polar H10} is worn-chest strap wearable measuring the HR via electrodes attached on a participant's chest \cite{polar_usa}.
    \item \textbf{Shimmer3 GSR+} measures GSR and the PPG using electrodes that are attached to the fingers \cite{shimmer}.
\end{itemize}

The behavioral sensors included in the monitoring system are as follows:
\begin{itemize}
    \item \textbf{Intel RealSense} is used to record 3D-depth and 2D color videos, and mounted on the top of the TV screen for capturing participant's face \cite{intel_realsense}. 
    \item\textbf{USB camera} is a basic camera to monitor the side view of the participants.
    \item \textbf{Mouse \& Keyboard} is used to track mouse cursor and monitor pushed keys.
    \item \textbf{Microphone } is used to record the participant's voice.
\end{itemize}

\subsection{Stimulus}
For the emotion elicitation task, the images and the audio clips were taken from the IAPS and IADS which are widely used and validated in the field of physiology for provoking specific emotions \cite{sanchez2018artificial, hsu2018affective}.
We particularly exploited 21 pictures of the International Affective Picture System (IAPS) \cite{lang1997international} and 21 audio clips of the International Affective Digitized Sound System (IADS) \cite{bradley1999international}. We used these visual and auditory stimuli to elicit targeted seven-emotions (e.g., happiness, sadness, anger, fear, surprise, and neutral). Table \ref{tab:used_datsets} shows the finally selected stimulus data for this user study. The used images and the number of IAPS and IADS are included on the dataset. For the cognitive workload task, we employed dual $N$-back games \cite{hampson2006brain}. 
To provoke different levels of cognitive workload (e.g., low, medium, and high), we controlled the number of back steps ($N$) of games from 1-back to 3-back to adjust the difficulty of the games. 


\begin{table}[]
\centering
    \caption{Selected stimulus data for basic emotions.}
    \label{tab:used_datsets}
    \resizebox{\columnwidth}{!}{
    \begin{tabular}{ccc}
        \hline        \hline

        \textbf{Type of emotion } & \textbf{IAPS images}   & \textbf{IADS: audios}       \\ \hline
        Happiness             & \#1710, \#2070, \#2550  & \#110, \#226, \#820 \\ \hline
        Sadness               & \#2800, \#3230, \#3350 & \#105, \#278, \#812 \\ \hline
        Anger                 & \#4621, \#6560, \#6840 & \#106, \#290, \#420 \\ \hline
        Fear                  & \#1120, \#1201, \#1930 & \#276, \#286, \#712 \\ \hline
        Surprise              & \#1616, \#3022, \#8180 & \#114, \#360, \#425 \\ \hline
        Disgust               & \#7380, \#9300, \#9320 & \#210, \#255, \#700 \\ \hline
        Neutral               & \#7080, \#7175, \#7217 & \#262, \#319, \#723 \\ \hline        \hline

    \end{tabular}
    }
\end{table}


\subsection{Experimental Protocol}
\begin{figure*}[t]
    \centering
     \begin{subfigure}[b]{0.315\linewidth}
        \centering 
        \includegraphics[width=1\linewidth]{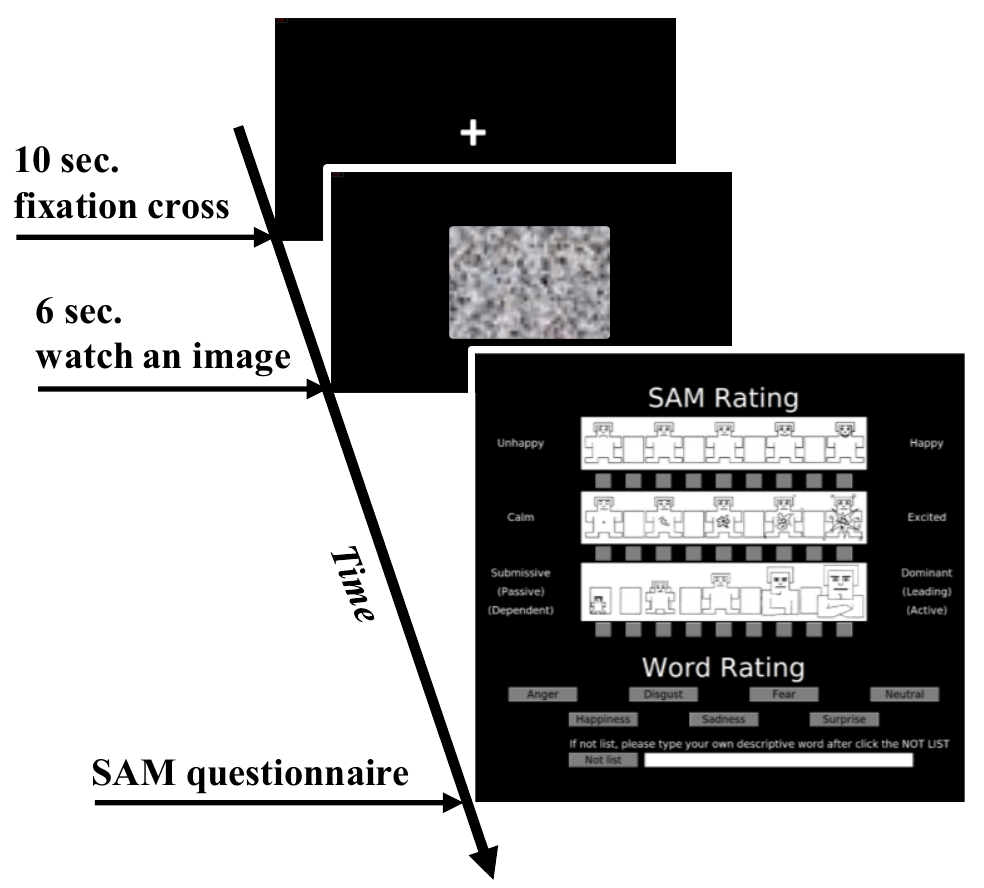} 
        \caption{}
        \label{img:emotion_test_image}
    \end{subfigure}
    \begin{subfigure}[b]{0.315\linewidth}
        \centering 
        \includegraphics[width=1\linewidth]{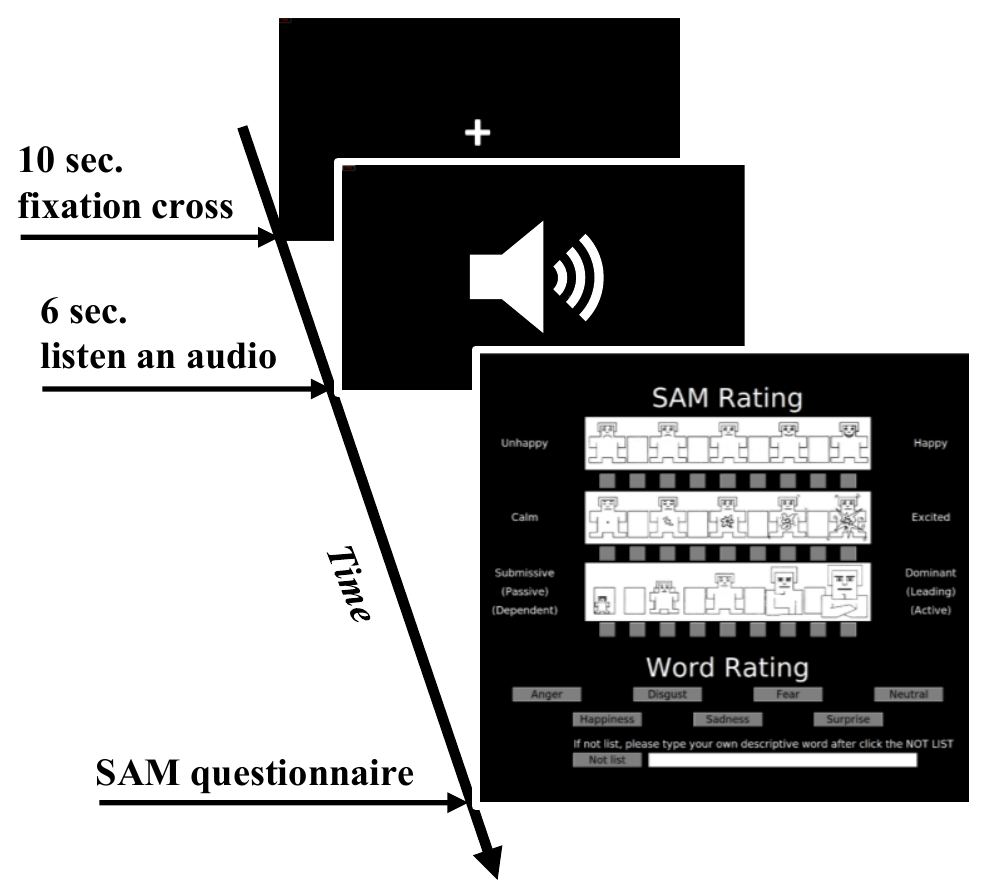} 
        \caption{}
        \label{img:emotion_test_audio}
    \end{subfigure}
    \begin{subfigure}[b]{0.35\linewidth}
        \centering 
        \includegraphics[width=1\linewidth]{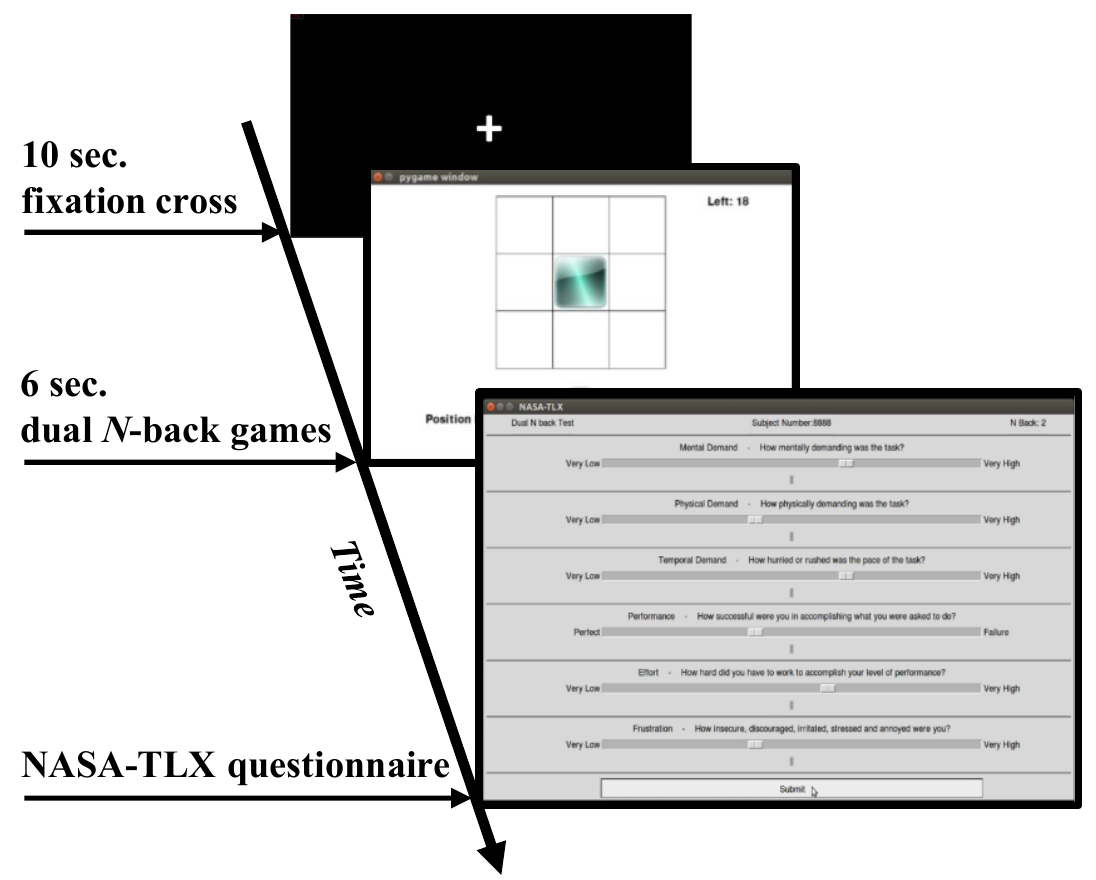} 
        \caption{}
        \label{img:workload_test_hor}
    \end{subfigure}
    \caption{Details of the procedures for emotion elicitation tasks and cognitive workload tasks in the user study; (a) using images of IAPS set, (b) using audio clips of IADS set, and (c) using a dual $N$-back game.}
    \label{img:detail_procedures}
\end{figure*}
In the user study, participants were given three tasks as illustrated in Fig. \ref{img:detail_procedures}. The first and second tasks are for emotional elicitation using IAPS and IADS, respectively. The third task is to stimulate the three-levels cognitive workload using dual $N$-back game. After finishing each task, the participant took a break until they want to proceed with the next task.


The first and second tasks were the emotion elicitation task which was composed of 21 rounds for each task. The participants were asked to look at a white cross on the screen for 10 seconds (called a fixation cross), then watch images of IAPS for 6 seconds in the first tasks or listen to short audio clips of IADS for 6 seconds in the second tasks, and then rate their perceived emotion with a 9-point Self-Assessment Manikin (SAM) scale \cite{bradley1994measuring}. The images and the audios were selected such that they can stimulate various human emotions. Fig. \ref{img:emotion_test_image} and Fig.\ref{img:emotion_test_audio} explain the procedures of the emotion elicitation task using the images and sound stimulus, respectively. 

The third task was the cognitive workload task which consisted of three rounds by presenting different levels of difficulty, low, medium, and high. The participants were asked to complete the Dual $N$-back games. 
During the experiment, the humans' physiological and behavioral conditions were monitored using the proposed monitoring system in section \ref{sec:data_collection_method}. After they completed each session, they were asked to rate their perceived cognitive workload with NASA-Task Load Index (NASA-TLX) \cite{hart1988development}. Fig. \ref{img:workload_test_hor} shows the procedures of the cognitive workload tasks.

\section{Dataset Construction}
\label{sec:data_construction}
In this section, we explain the details of the proposed dataset configuration: physiological and behavior sensor data. 
Table \ref{tab:dataset_summary} presents the summary of the dataset.

\begin{table}[t]
    \caption{Summary of the Dataset.}
    \label{tab:dataset_summary}
    \centering
    \resizebox{\columnwidth}{!}{
    \begin{tabular}{ll}
    \hline    \hline

    \textbf{Participants} & 30 (Female: 11 and Male: 19)                                      \\\hline
    \textbf{Number of ROSbag files} & 1,602 files (about 787 GB)                     \\\hline
    \textbf{Emotion ratings}        & \begin{tabular}[c]{@{}l@{}}Arousal, Valence, Dominance, \\Word-emotion rating\end{tabular} \\\hline
    \textbf{Workload rating} &
      \begin{tabular}[c]{@{}l@{}}Mental/Physical/Temporal demand, \\Performance, Effort, Frustration \end{tabular} \\\hline
    \textbf{Physiological signals} &
      \begin{tabular}[c]{@{}l@{}}PPG from wist and chest, EDA, IBI, \\ ST, ECG, GSR, and EMG  \end{tabular} \\\hline
    \textbf{Video types}            & \begin{tabular}[c]{@{}l@{}}Frontal face videos (RGB and depth), \\ side view video\end{tabular}                \\ \hline\hline
    \end{tabular}
    }
\end{table}
\subsection{Physiological Sensor Data} 
The dataset includes BVP, ST, EDA, and IBI from Empatica E4 sensor with 30Hz sampling time, BVP and GSR from Shimmer3 GSR unit with 30Hz sampling time, HR from Polar H10 with 1Hz, and 8-channel EMGs from Myo armband with 50 sampling time.

\begin{figure}[t]
    \centering
        \includegraphics[width=1\linewidth]{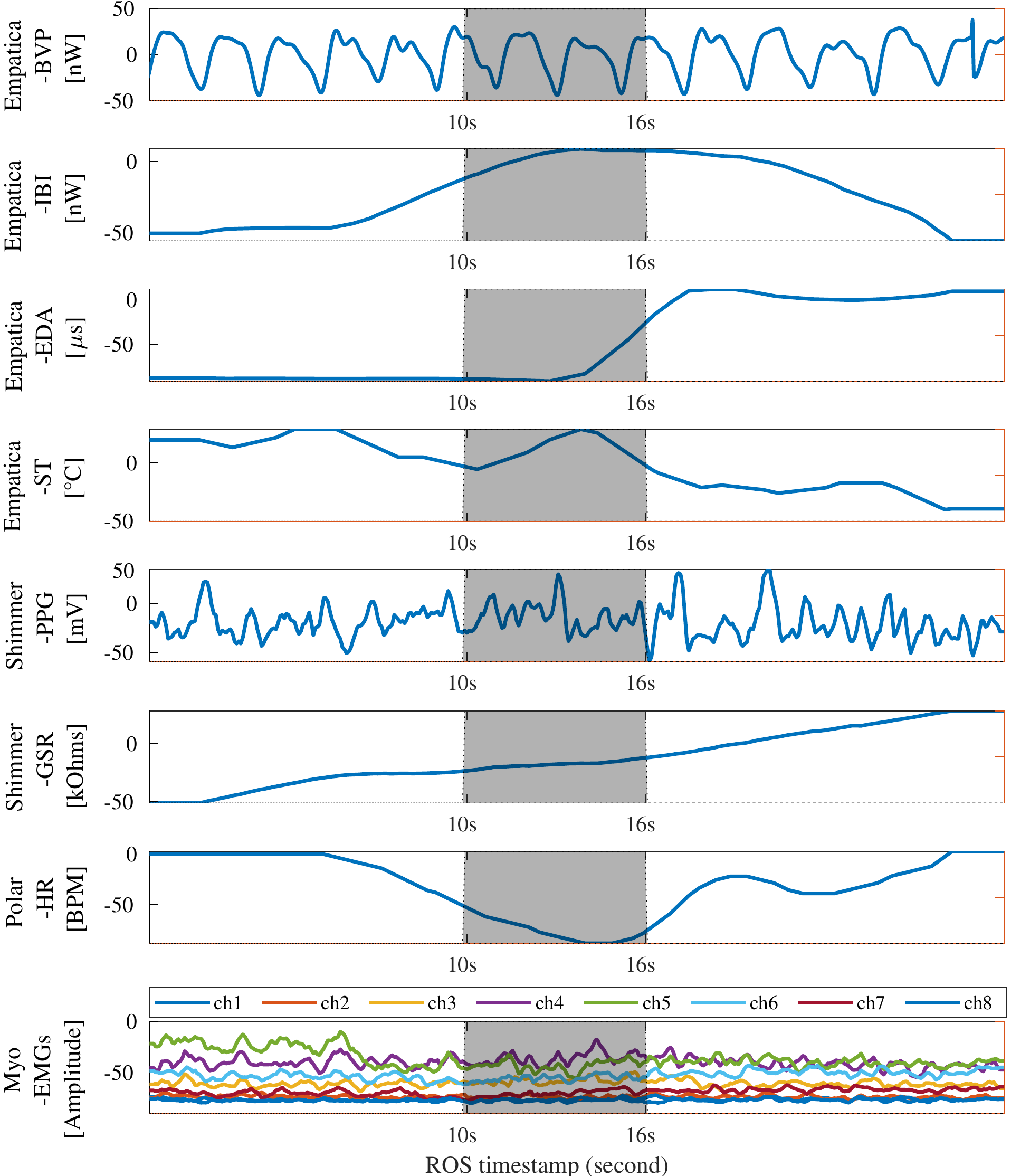} 
    \caption{Example of the physiological signals from the dataset (IAPS\#1201 of P13); IBI, EDA, and ST data of Empatica E4, PPG and GSR data of Shimmer, HR of Polar H10, and 8-channels EMGs of Myo armband (from top to bottom).}
    \label{img:example_physiology_data}
\end{figure}

Fig. \ref{img:example_physiology_data} shows an example of physiological data in the dataset (IAPS \#1201, P13).
The first plot from top is the BVP signals, the second plot is the average of the IBI data, the third plot is the average of the EDA, the fourth plot is the average of ST data. Those data are collected from the Empatica E4 sensor. The fifth and sixth plots are raw PPG and GSR data of the Shimmer3 sensor. The seventh plot is the result of HR data of the Polar H10. The last plot is raw data of 8-channel EMGs of the Myo armband.

In the figures, the gray area indicates the duration when the stimulus was exposed to the participants during the experiments.
The left side of the gray area is a baseline section where the participant lies in the fixation section. The right side of the gray area is a self-assessment reporting section for participants to fill the subjective questionnaires out.

Table \ref{tab:physiology_rostopic_list} summarizes the rostopic message information of the physiological data in the dataset.

\begin{table}[h]
\caption{List of rostopic messages for physiological sensors.}
    \label{tab:physiology_rostopic_list}
\resizebox{\columnwidth}{!}{
\begin{tabular}{cll}
\hline\hline
\textbf{Sensor} & \multicolumn{1}{c}{\textbf{
Name of rostopic messages}} & \multicolumn{1}{c}{\textbf{Type of rostopic messages}} \\ \hline
Empatica E4                  & \textit{/physiological\_data}  & empatica\_e4\_msgs/DataArrays\footnote{Empatica E4 ROS message: \url{https://github.com/hyeonukbhin/empatica_e4_msgs}}\\ \hline
\multirow{2}{*}{Shimmer3}    & \textit{/shimmer3/GSR}      & std\_msgs/Float64             \\
                             & \textit{/shimmer3/PPG}      & std\_msgs/Float64             \\ \hline
Polar H10                    & \textit{/polar\_h10/hrv}    & std\_msgs/Int32MultiArray     \\ \hline
\multirow{2}{*}{Myo Armband} & \textit{/myo\_raw/myo\_emg} & ros\_myo/EmgArray\footnote{Myo Armband ROS message: \url{https://github.com/dzhu/myo-raw}}\\                             & \textit{/myo\_raw/myo\_imu} & sensor\_msgs/Imu               \\ \hline
\end{tabular}
}
\end{table}

\subsection{Behavioral Sensor Data}

The dataset includes three different kinds of image sequences taken by two cameras. The Intel RealSense camera located at the front captured facial expressions and upper body gestures in 30 frames per second (fps). At the same time, depth camera results separately were recorded in 30 fps. The USB camera at the side of participants obtained induced behavioral responses in 10 fps. As well, the participant's speech was recorded via a microphone mounted on the participant's neck for the user study.

\begin{figure} 
    \centering
    \begin{subfigure}[b]{1\linewidth}
        \centering 
            \includegraphics[width=1\linewidth]{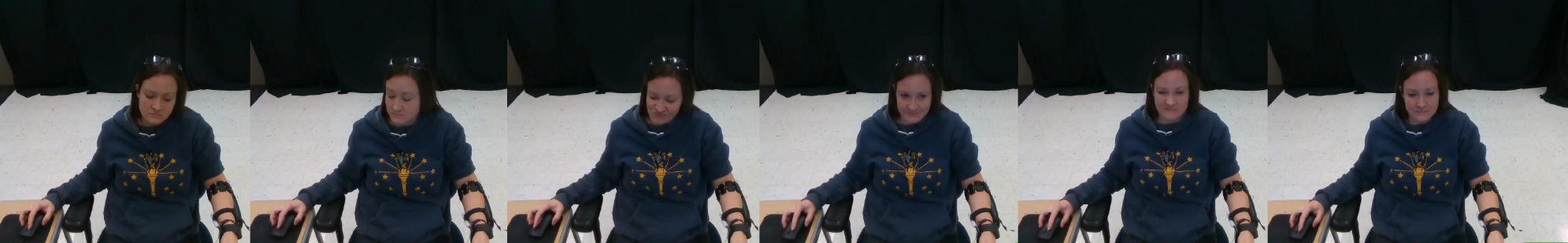} 
        \caption{Front view sequence; 8, 11, 14, 17, 20, 23 seconds } 
        \vspace{3pt}        
        \label{img:behavior_emotion_frontview}
    \end{subfigure}
    \begin{subfigure}[b]{1\linewidth}
        \centering 
        \includegraphics[width=1\linewidth]{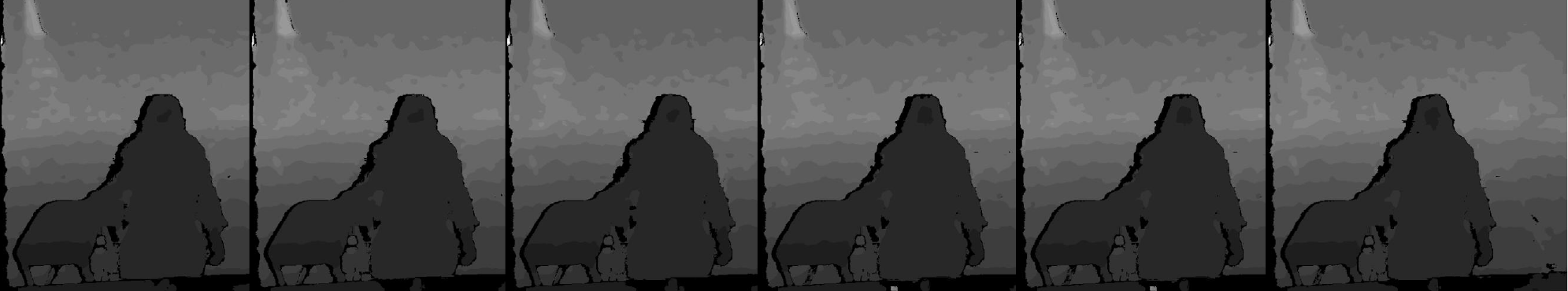} 
        \caption{Depth view sequence}
\vspace{3pt}        
        \label{img:behavior_emotion_depthview}
    \end{subfigure}
    \begin{subfigure}[b]{1\linewidth}
        \centering 
        \includegraphics[width=1\linewidth]{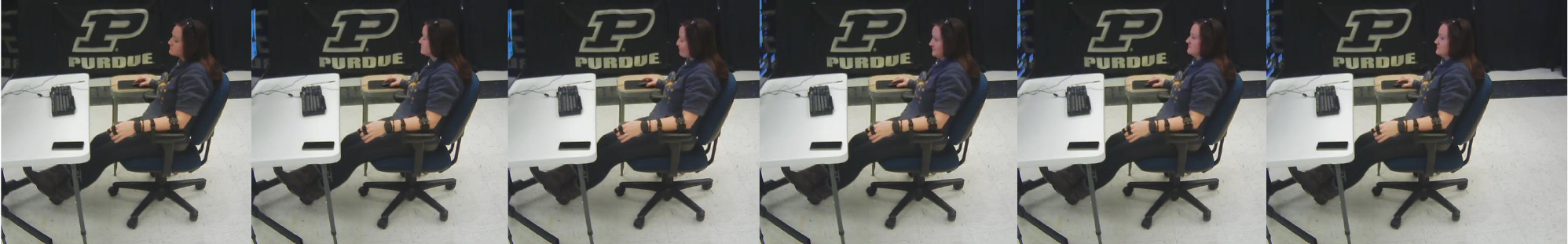} 
        \caption{Side view sequence}
        \label{img:behavior_emotion_sideview}
    \end{subfigure}
    \begin{subfigure}[b]{1\linewidth}
        \centering 
        \includegraphics[width=1\linewidth]{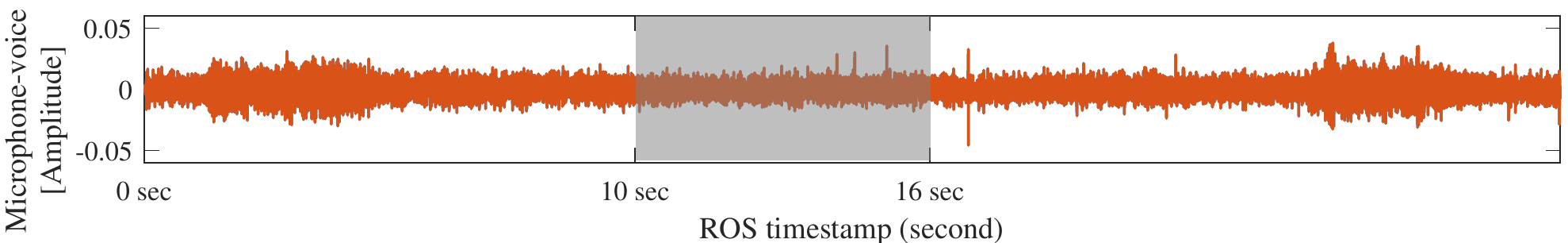} 
        \caption{Speech signals}
        \label{img:speech_signal_plot}
    \end{subfigure}
    \begin{subfigure}[b]{1\linewidth}
        \centering 
        \includegraphics[width=1\linewidth]{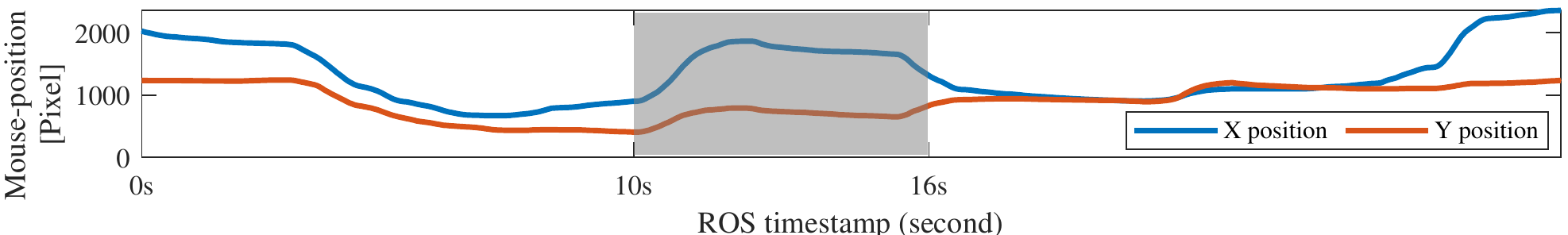} 
        \caption{Positions of the mouse cursor}
        \label{img:mouse_position_plot}
    \end{subfigure}
    
    \caption{Example of the behavioral data from the dataset (IAPS\#1201 of P13); (a) Front RGB images, (b) Front depth images, (c) Side-view images, (d) speech signals, and (e) positions of the mouse cursor.
    }
    \label{img:example_emotion_elicitation}
\end{figure}

The collected experimental data showed that the tasks elicited participants' emotional and cognitive states. For example, a piece of the proposed dataset with the participant P13 and visual stimulus IAPS\#1201 is shown in Fig. \ref{img:example_emotion_elicitation}. Given the recorded stream of participants, as presented in Fig. \ref{img:behavior_emotion_frontview}, \ref{img:behavior_emotion_depthview} and \ref{img:behavior_emotion_sideview}, the behavioral data include facial expressions and body movements, which may imply emotional reactions. 

Table \ref{tab:behavior_rostopic_list} summarizes the rostopic messages information of the behavioral data in the dataset.

\begin{table}[h]
\caption{List of rostopic messages for behavioral sensors.}
    \label{tab:behavior_rostopic_list}
\resizebox{\columnwidth}{!}
{
\begin{tabular}{cll}
\hline\hline
\textbf{Devices}                                                           & \multicolumn{1}{c}{\textbf{Name of rostopic message}} & \multicolumn{1}{c}{\textbf{Type of rostopic message}} \\ \hline
\multirow{2}{*}{Intel RealSense} & \textit{/camera/color/image\_raw}           & sensor\_msgs/Image                             \\
                       & \textit{/camera/depth/image\_rect\_raw} & sensor\_msgs/Image        \\ \hline
USB caemra                       & \textit{/image\_raw}                        & sensor\_msgs/Image                             \\ \hline
Microphone             & \textit{/audio/audio}                   & audio\_common\_msgs/AudioData       \\ \hline
\multirow{2}{*}{Mouse} & \textit{/mouse\_tracking/click}         & std\_msgs/String          \\
                       & \textit{/mouse\_tracking/position}      & std\_msgs/Int32MultiArray \\ \hline
Keyboard               & \textit{/keyboard\_tracking/info}       & std\_msgs/String          \\ \hline\hline
\end{tabular}
}

\end{table}

\section{Subjective Rating Analysis} 
\label{sec:rating_analysis}
\subsection{The SAM Rating in the Emotion Elicitation Task}
\begin{figure}
    \centering
     \begin{subfigure}[b]{1\linewidth}
        \centering 
        \includegraphics[width=1\linewidth]{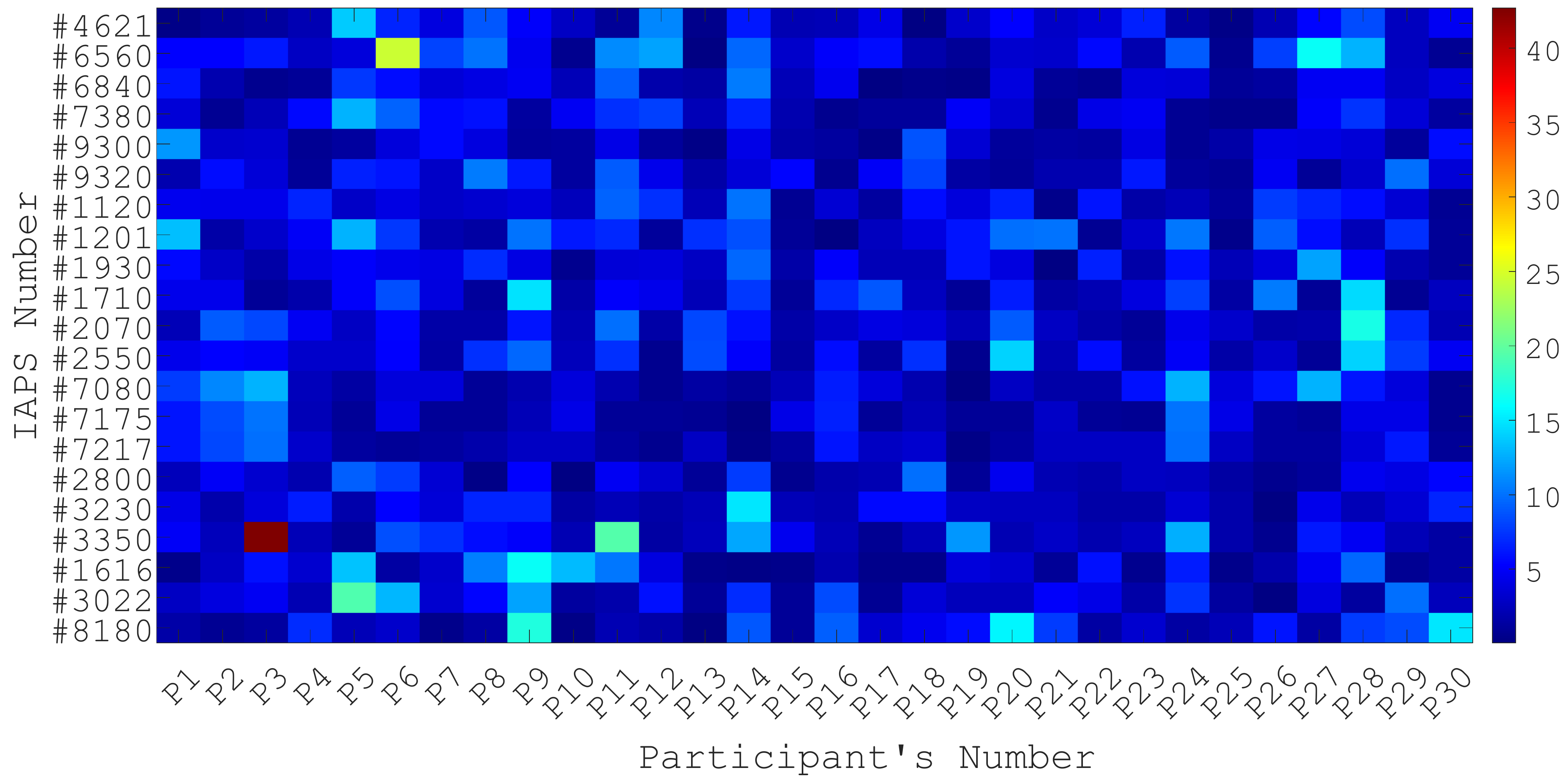} 
        \caption{Results of RMSE between each participant and IAPS sets.}
        \label{img:sam_compare_iaps}
    \end{subfigure}
    
    \begin{subfigure}[b]{1\linewidth}
        \centering 
        \includegraphics[width=1\linewidth]{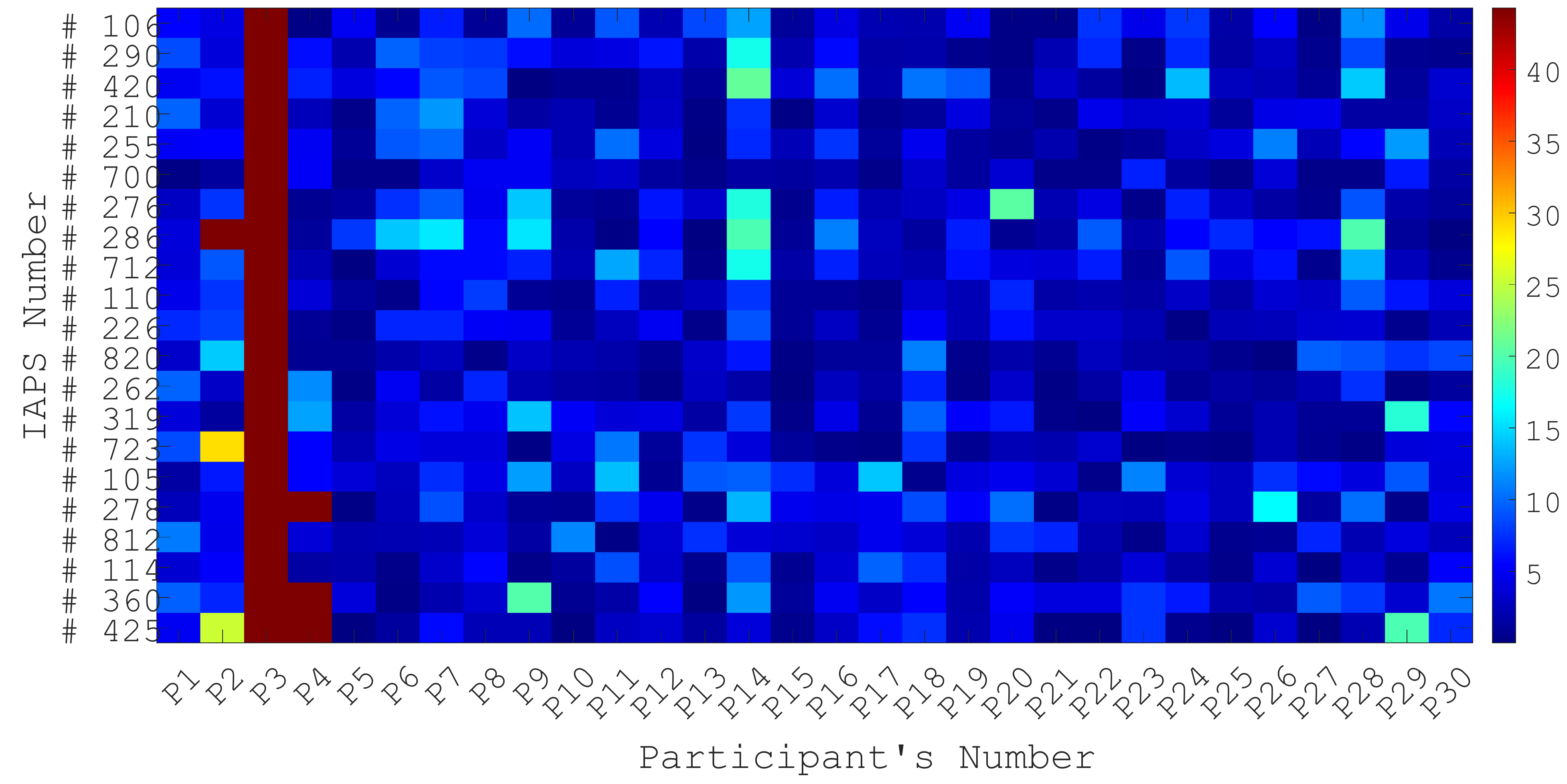} 
        \caption{Results of RMSE between each participant and IADS sets.}
        \label{img:sam_compare_iads}
    \end{subfigure}
    
    \caption{Color map to display the comparing results of root mean square error (RMSE) between the collected SAM rating in the emotion elicitation tasks and the reference rating of (a) IAPS and (b) IADS set.} 
    \label{img:sam_result}
\end{figure}

All participants' subjective measures (e.g., arousal, valence, and dominance) in each emotion elicitation task are compared to the reference values published in \cite{lang2008technical,bradley2007international}. The comparison results were plotted on a grid map image as shown in Fig. \ref{img:sam_result}, where we used Root-Mean Square Error (RMSE). 
Fig. \ref{img:sam_compare_iaps} shows the result of the comparison analysis in the emotion elicitation task using IAPS.
Fig. \ref{img:sam_compare_iads} shows the result of the comparison analysis in the emotion elicitation task using IADS.
In both figures, the x-axis is the participant's number from P1 to P30 and the y-axis is the number of the dataset. 
In order to show the overall results of the comparison analysis of the self-assessments, we displayed the results using gradual colors from blue to red. 
The closer the index value to 0 (blue) means that the more similar it is to the reference value. On the other hand, the closer the index value to 45 (red) means that the more different it is is from the reference value.

For the results of the SAM scales in the emotion elicitation task using IAPS, the lowest similarity of the dataset is \#3350 of P3 with RMSE $42.69$, and the highest similarity of the dataset is IAPS\#3022 of P26 with RMSE $0.04$. 
P25 produced the highest similarity with mean RMSE $1.66$, and P28 produced the lowest similarity with mean RMSE $6.81$. The overall average and standard deviation of RMSE are $4.26$ and $3.90$, respectively.

For the SAM scales in the emotion elicitation task using IADS, the lowest quality of the dataset is IADS\#286 of P2 with RMSE $44.28$, and the highest quality of the dataset is \# 820 of P26 with RMSE $0.01$. 
P15 produced the highest similarity with mean RMSE $1.69$, and P14 produced the lowest similarity with mean RMSE $10.02$. The overall average and standard deviation of RMSE are $4.31$ and $4.41$, respectively, excepting lost data (P3's data and P4's \#278, \#360, and \#425).

\subsection{NASA-TLX Rating in the Cognitive Workload Task: }
\begin{figure}[t]
    \centering
        \includegraphics[width=1\linewidth]{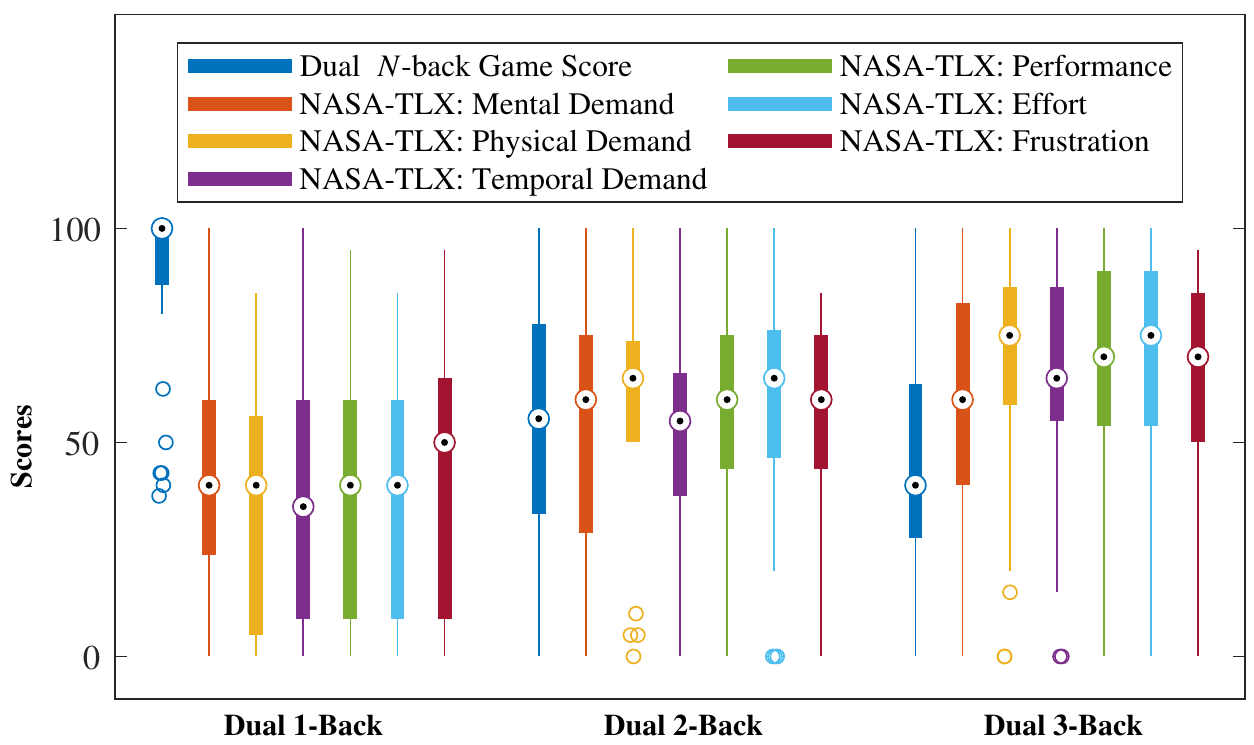} 
    \caption{The results of the Dual $N$-Back game score and NASA-TLX questionnaire according to the level of the workload.}
    \label{img:nasa_tlx_result}
\end{figure}

We analyzed the results of the NASA-TLX rating scales and scores of the dual $N$-back game to monitor the change of the participant's workload.
Fig. \ref{img:nasa_tlx_result} shows the overall results of the NASA-TLX and dual $N$-back game. The blue bar means the score of the dual $N$-back game, and orange, yellow, purple, green, sky-blue, and red bars mean each subscale ratings of the NASA-TLX: mental demand, physical demand, temporal demand, overall performance, effort, and frustration level that are rated within a 100-points range.
In the dual 1-back, most participants obtained 100 point scores in the dual $N$-back game, and also acquired the lowest rating of the subscales in the NASA-TLX (median of each the subscales: 40/40/35/40/40/50).
In the dual 2-back, the participants' game scores decreased $55.56$ points compared to the result of the dual 1-back. On the other hand, the subscales of the NASA-TLX increased; (median: 60/65/55/60/65/60). In the dual 3-back, the game score is 40-points that is the lowest score and all subscales of the NASA-TLX are highest scores compared to others; (median: 70/75/65/70/75/70).


\section{Access to Dataset and Application}
\label{sec:access_dataset}
To obtain the permission for accessing the dataset presented in this paper, researchers should contact us via email; \href{mailto:info@smart-laboratory.org}{info@smart-laboratory.org}. We will also provide source codes (such as ROS package and Matlab codes) to replay the dataset. 
\subsection{Examples of replaying the dataset}
Since the dataset is encapsulated into the ROSbag files, the dataset can be easily played back in in any ROS-compatible robot system, such as ROS system in Linux system, and Matlab.

For using the ROS system, users should install the ROS on Linux, and then decompress the compressed dataset. An example of reading a ROSbag file on Linux system is below:

\begin{footnotesize}
\begin{lstlisting}
  $ rosbag decompress [rosbag_name.bag]
  $ rqt_bag or rosbag play [rosbag_name.bag]
\end{lstlisting}
\end{footnotesize}

For using Matlab, users should install ROS toolbox\footnote{ROS Toolbox in Matlab: \url{https://www.mathworks.com/products/ros.html}} that enables accessing the ROS and exchanging data. An example of reading a ROSbag file in Matlab is below:

\lstset{
  style              = Matlab-editor,
  basicstyle         = \mlttfamily\footnotesize,
  escapechar         = ",
  mlshowsectionrules = true,
}
\begin{lstlisting}
  % Read a rosbag file
  input_bag = rosbag('[rosbag_name].bag');
  % Display availiable topics included in the rosbag
  input_bag.AvailableTopics
  % Display all message data along with time stamps.
  input_bag.MessageList;
  % Select topics from the all message list
  selected_topic = select(input_bag, 'Topic', '[rostopic name]]');
  selected_topic_msgStructs= readMessages(selected_topic, 'DataFormat', 'struct');
\end{lstlisting}

\subsection{Applications: Facial Emotion Analysis}
In this subsection, we demonstrate an example of applications using the proposed affective dataset. The application is to estimate human facial expression from the facial video of the affective dataset using open source-based Face Emotion Recognition (FER) libraries \cite{pypi}.

\begin{figure}
    \centering
    \begin{subfigure}[b]{1\linewidth}
        \centering 
        \includegraphics[width=1\linewidth]{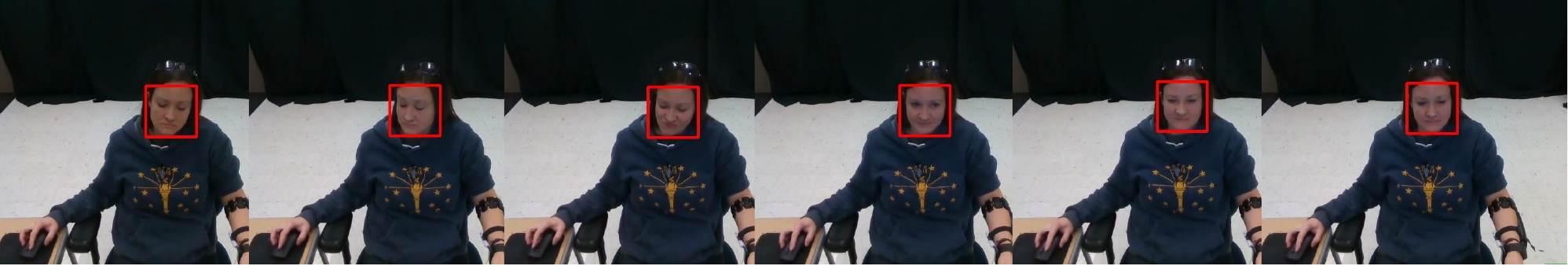} 
        \caption{Facial expression images of Participant \#13}
        \label{img:behavior_emotion_front}
    \end{subfigure}

\vspace{4pt} 
    \begin{subfigure}[b]{1\linewidth}
        \centering 
        \includegraphics[width=1\linewidth]{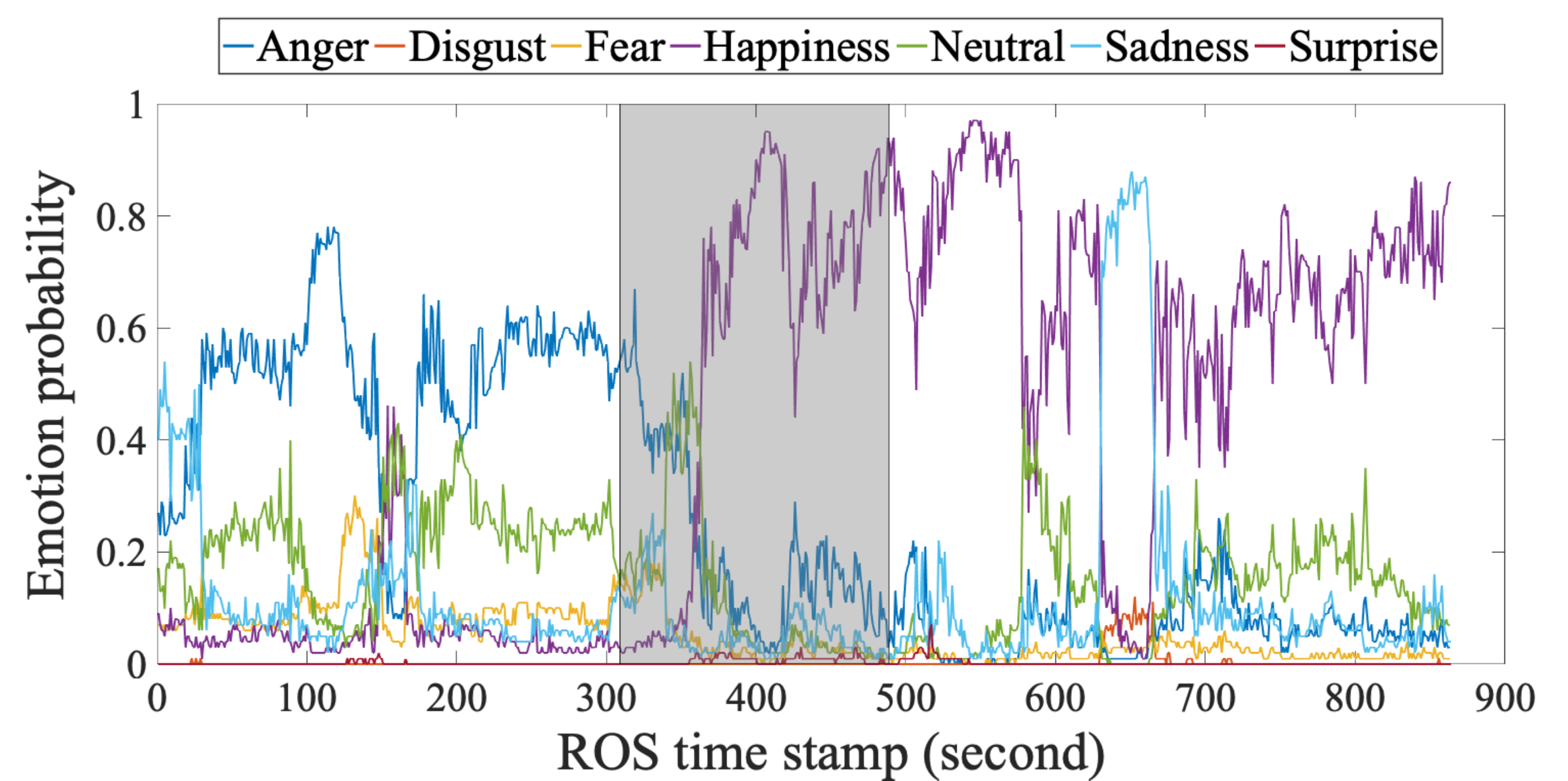} 
        \caption{Emotion probability from user study}
        \label{img:behavior_emotion_probability}
    \end{subfigure}
    \caption{An example of emotion analysis from the collected facial expression images.}
    \label{img:application_example}
\end{figure}

Fig. \ref{img:behavior_emotion_probability} shows computed emotion based on the facial expressions (P13's IAPS\#1201). The gray area in Fig. \ref{img:behavior_emotion_probability} indicates the exposure duration of visual or auditory stimuli. The left side of the gray area is the exposure time with 10 seconds fixation cross. The right side of the gray area indicates the period during the self-assessment. 
The participant P13 rated the emotion response as word-emotion rating `Disgust' with the SAM scale assessment level. Compared to the highest emotion probability of `Happiness' from the emotion recognition library in Fig. \ref{img:behavior_emotion_probability}, not only is the calculated emotion different from self-assessed one, but the facial expressions are also not matched with the SAM scale assessment. This implies that only analyzing facial expressions may not be enough to fully estimate human emotions and that other behavioral or physiological features and analysis may need to be combined.

\section{Conclusion and Future Works}
\label{sec:conclusion_future_work}
In this paper, we have introduced a new ROSbag-based affective dataset that shows how one's emotional and cognitive states affect  physiological and behavioral data.

For building the affective dataset, we designed a user study to stimulate the targeted emotions using IAPS and IADS datasets and different levels of the cognitive workload using dual $N$-back games, and executed the study by recruiting 30 participants. In the user study, we recorded the particiapnts status that includes physiological data from commercial wearable devices and the behavioral data using hardware devices, as well as the results of the subjective questionnaires using SAM and NASA-TLX. All data were saved in single ROSbag files rather than CSV files. This not only ensures synchronization of signals and videos in the dataset, but also allows researchers to easily analyze and verify their algorithms by connecting directly to this dataset through ROS. The generated dataset consists of 1,602 ROSbag files, and the size of the dataset is about 787GB. We expect that our dataset can be a great resource for many researchers in the fields of affective computing, HCI, and HRI.

In the future, we will utilize more (and latest) physiological sensors and hardware devices and develop additional psychological experiments related to workload, in order to update the affective dataset. We also plan to analyze more details of the dataset by extracting features from the collected data and validate the dataset using advanced machine learning techniques to estimate human's emotional and cognitive states.

\addtolength{\textheight}{-12cm}   


\section*{Acknowledgment}
This material is based upon work supported by the National Science Foundation under Grant No. IIS-1846221.

\bibliography{root}
\bibliographystyle{IEEEtran}
\end{document}